\documentclass[12pt]{article}
\usepackage{graphicx}

\def\d{\delta}

\def\g{\gamma}

\def\l{\lambda}
\def\m{\mu}
\def\n{\nu}

\def\p{\pi}

\def\s{\sigma}

\def\G{\Gamma}

\def\L{\Lambda}

\def\R{\mathcal R}

\def\beq{\begin{equation}}
\def\eeq{\end{equation}}
\def\bea{\begin{eqnarray}}
\def\eea{\end{eqnarray}}

\def\NO{\nonumber}

\def\pl#1#2#3{Phys.~Lett.~{\bf B {#1}} ({#2}) #3}

\def\prl#1#2#3{Phys.~Rev.~Lett.~{\bf #1} ({#2}) #3}
\def\pr#1#2#3{Phys.~Rev.~{\bf D {#1}} ({#2}) #3}

\def\ap#1#2#3{Ann.~of Phys.~{\bf {#1}} ({#2}) #3}
\def\prep#1#2#3{Phys.~Rep.~{\bf {#1}C} ({#2}) #3}
\def\ptp#1#2#3{Progr.~Theor.~Phys.~{\bf {#1}} ({#2}) #3}

\renewcommand{\(}{\left(}
\renewcommand{\)}{\right)}

\newcommand{\tr}{\mbox{tr}}

\begin{document}
\date{\mbox{ }}
\title{{\normalsize DESY 01-106\hfill\mbox{}\\
August 2001\hfill\mbox{}}\\
\vspace{2cm} \textbf{Gauge Unification In Six Dimensions}\\
[8mm]}
\author{T.~Asaka, W.~Buchm\"uller, L.~Covi\\
\textit{Deutsches Elektronen-Synchrotron DESY, Hamburg, Germany}}
\maketitle

\thispagestyle{empty}

\begin{abstract}
\noindent
We study the breaking of a supersymmetric SO(10) GUT in 6 dimensions by 
orbifold compactification. In 4 dimensions we obtain a N=1 supersymmetric 
theory with the standard model gauge group enlarged by an additional U(1)
symmetry. The 4-dimensional gauge symmetry is
obtained as intersection of the Pati-Salam and the Georgi-Glashow subgroups
of SO(10), which appear as unbroken subgroups in the two 5 dimensional
subspaces, respectively. The doublet-triplet splitting arises  as in the 
recently discussed SU(5) GUTs in 5 dimensions.  
\end{abstract}

\newpage
The simplest grand unified theory (GUT) which unifies one generation of
quarks and leptons, including the right-handed neutrino, in a single
irreducible representation is based on the gauge group SO(10)\cite{gfm75}.
Important subgroups, whose phenomenology has been studied in great detail,
are the Pati-Salam group SU(4)$\times$SU(2)$\times$SU(2) \cite{ps74} and 
the GUT group SU(5) of Georgi and Glashow \cite{gg74}.

The breaking of these GUT groups down to the standard model gauge group is
in general rather involved and requires often large Higgs representations.
In particular, the mass splitting between the weak doublet and the
colour triplet Higgs fields requires either a fine-tuning of parameters or 
additional, sophisticated mechanisms. An attractive new possibility has
been suggested by Kawamura \cite{kaw00}: compactifying a SU(5) GUT in 
5 dimensions (5d) on an orbifold, which breaks SU(5) to the standard model 
group SU(3)$\times$SU(2)$\times$U(1) only on one of the boundary branes, one 
achieves in 4d the correct gauge symmetry breaking and in addition the
wanted doublet-triplet splitting. Various aspects of such 5d SU(5) GUTs, 
including fermion masses, have recently been studied by several groups
\cite{afx01,hmr01}. The goal of the present paper is to extend the orbifold
breaking to the GUT group SO(10).

The breaking of G=SU(5) is achieved by means of a `parity' $P$, under which 
the generators of G are either even (S) or odd (A),
\bea
P S P^{-1} = S \;,\quad P A P^{-1} = -A\;.
\eea
Clearly, the set of generators S and A satisfy the relations
\bea
[S,S] \subseteq S\;,\quad [S,A] \subseteq A\;,\quad
[A,A] \subseteq S\;.
\eea
Hence, the group G$_S$ generated by $S$ is a symmetric subgroup of G.
Together with the identity $P$ forms the discrete group $Z_2$. In 
orbifold compactifications of 5d SU(5) GUTs the theory is assumed to
be invariant under the parity transformation of gauge fields $V^M(x,y)$,
$M = (\m,5)$, $\m=0\ldots 3$, $x^5=y$, and matter fields $H_{\pm}(x,y)$,
\bea\label{trafo}
P V_\mu(x,-y+a) P^{-1} = + V_\mu(x,y+a)\;,
\quad  P V_5 (x,-y+a) P^{-1} = - V_5 (x,y+a)\;,
\eea
\bea
P H_{\pm}(x,-y+a) = \pm H_{\pm}(x,y+a)\;,
\eea
where $a = 0$ or $a=\pi R/2$, and $R$ is the radius of the compact dimension. 
On the orbifold, $M=\R^4\times S^1/Z_2$, the GUT symmetry is then broken 
on branes which are fixed points of the transformation. Note, that this 
symmetry breaking preserves the rank of the group\footnote{For a detailed
discussion of GUT breaking by orbifolding, see e.g. \cite{hmr01}.}.

The extension of this procedure to the GUT group SO(10) is not straightforward
since the standard model group G$_{SM}$ is not a symmetric subgroup 
of SO(10)\cite{sla81,hmr01}. It is remarkable, however, that the extended 
standard model group G$_{SM'}$=G$_{SM}$$\times$U(1) is the maximal common 
subgroup of two symmetric subgroups of SO(10), SU(5)$\times$U(1) and
SU(4)$\times$SU(2)$\times$SU(2) (cf. fig. 1). This suggests to realize
the wanted SO(10) breaking by starting in 6 dimensions and orbifolding to
the two different subgroups in the two orthogonal compact dimensions.
\begin{figure}
\centering 
\includegraphics[scale=0.35]{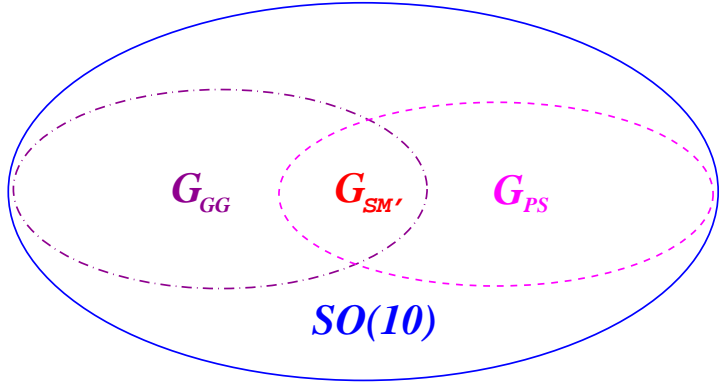}
\caption{The extended standard model gauge group 
G$_{SM'}$=SU(3)$\times$SU(2)$\times$U(1)$^2$ as intersection of the two
symmetric subgroups of SO(10), 
G$_{GG}$ = SU(5)$\times$U(1) and 
G$_{PS}$~=~SU(4)$\times$SU(2)$\times$SU(2).}
\end{figure}

Such an extension to 6 dimensions is also welcome with respect to 
supersymmetry. It is well known that the extended supersymmetry in 4d
is easily obtained by dimensional reduction from 6d \cite{fay85}. The 
corresponding unwanted N=2 supersymmetry in 4d can also be broken by 
appropriate orbifolding \cite{mp98}. The irreducible N=1 multiplets in 6d  
then split into reducible N=1 multiplets in 4d.

In the following we shall start from an SO(10) GUT in 6 dimensions. First,
we consider the compactification on a torus, i.e., $M=\R^4\times T^2$.
The goal is the construction of a N=1 supersymmetric Yang-Mills theory with
extended standard model gauge symmetry. The necessary breaking of the
extended supersymmetry in 4d and the breaking of the SO(10) GUT group
are then achieved in three steps of orbifolding leading finally to the
manifold  $M=\R^4\times T^2/(Z_2\times Z_2^{GG}\times Z_2^{PS})$.

Let us then consider the N=1 supersymmetric Yang-Mills theory in 6 dimensions.
The corresponding lagrangian reads \cite{fay85},
\bea
{\cal L}_{6d}^{YM} = \tr\(-{1\over 2} V_{MN}V^{MN} 
+ i \bar{\Lambda}\G^M D_M \L\)\;.
\eea
Here $V_M = T^A V^A_M$ and $\L = T^A \L^A$, where $T^A$ are the SO(10)
generators; $D_M\L = \partial_M\L -ig[V_M,\L]$ and 
$V_{MN} = [D_M,D_N]/(ig)$; the $\G$-matrices are
\bea
\G^{\m} = \(\begin{tabular}{cc} $\g^{\m}$&0 \\ 0&$\g^{\m}$ \end{tabular}\)\;,
\quad
\G^{5} = \(\begin{tabular}{cc} 0&$i\g_5$ \\ $i\g_5$&0 \end{tabular}\)\;,
\quad
\G^{6} = \(\begin{tabular}{cc} 0&$\g_5$ \\ $-\g_5$&0 \end{tabular}\)\;,
\eea
with ${\g_5}^2=I$ and $\{\G_M,\G_N\}=2\eta_{MN}$=diag(1,-1,-1,-1,-1,-1). The
gaugino is composed of two Weyl fermions with opposite 4d chirality,
$\L = (\l_1,-i\l_2)$, with $\g_5 \l_1 = -\l_1$ and $\g_5 \l_2 = \l_2$.
$\L $ has negative 6d chirality, $\G_7 \L = -\L$, with 
$\G_7$ = diag($\g_5,-\g_5$).

We now perform a torus compactification, i.e., we choose the manifold
$M=\R^4\times T^2$. For the fields $\Phi = (V_M,\L)$ one then has the
mode expansion ($x^6=z$),
\bea
\Phi(x,y,z) = {1\over 2\pi\sqrt{R_1R_2}}\sum_{m,n}\Phi^{(m,n)}(x)
              \exp{\left\{i\({my\over R_1}+{nz\over R_2}\)\right\}}\;,
\eea
where $R_1$ and $R_2$ are the two radii of the torus. Since the vector
field is hermitian the corresponding coefficients satisfy the relation
$V_M^{(-m,-n)}= V_M^{(m,n)\dagger}$.  

It is straightforward to work out the 4d lagrangian obtained by integrating
over the 5th and the 6th dimensions. For the 4d scalars a convenient choice
of variables is
\bea
\Pi_1^{(m,n)}(x) &=& {i\over M(m,n)}
\({m\over R_1} V_5^{(m,n)}(x) + {n\over R_2} V_6^{(m,n)}(x)\)\;,\\
\Pi_2^{(m,n)}(x) &=& {i\over M(m,n)}
\(-{n\over R_2} V_5^{(m,n)}(x) + {m\over R_1} V_6^{(m,n)}(x)\)\;,
\eea
where $M(m,n) = \sqrt{\({m\over R_1}\)^2 + \({n\over R_2}\)^2}$.
The kinetic term for gauge fields and scalar fields is then given by
\bea\label{kkhiggs}
{\cal L}_4^{(1)} &=& \sum_{m,n} \tr\( 
-{1\over 2} \widetilde{V}_{\m\n}^{(m,n)\dagger}\widetilde{V}^{(m,n)\m\n}
+ M(m,n)^2 V_{\m}^{(m,n)\dagger}V^{(m,n)\m}\right.\NO\\
&&\hspace{1cm}+\partial_\m\Pi_2^{(m,n)\dagger}\partial^\m\Pi_2^{(m,n)}
+ M(m,n)^2 \Pi_2^{(m,n)\dagger}\Pi_2^{(m,n)}\NO\\
&&\hspace{1cm}+\partial_\m\Pi_1^{(m,n)\dagger}\partial^\m\Pi_1^{(m,n)}\NO\\
&&\hspace{1cm}\left. - M(m,n)
\(V_\m^{(m,n)\dagger} \partial^\m \Pi_1^{(m,n)}
+\partial^\m \Pi_1^{(m,n)\dagger}V_\m^{(m,n)}\)\)\;,
\eea
where $\widetilde{V}_{\m\n}^{(m,n)} = 
\partial_\m V_\n^{(m,n)} - \partial_\n V_\m^{(m,n)}$. Massless states are
obtained for $m=n=0$ (zero modes). The mass generation for the massive
Kaluza-Klein (KK) states is analogous to the Higgs mechanism. Here 
$\Pi_1^{(m,n)}$ play the role of the Nambu-Goldstone bosons and $M(m,n)$ 
correspond to the Higgs vacuum expectation values.  

Similarly, one obtains for the gauginos,
\bea
{\cal L}_4^{(2)} &=& \sum_{m,n} \tr\(
i\overline{\l_1}^{(m,n)}\g^\m\partial_\m \l_1^{(m,n)} +
i\overline{\l_2}^{(m,n)}\g^\m\partial_\m \l_2^{(m,n)}  \right.\NO\\
&& \hspace{1cm}\left. - \({m\over R_1} -i{n\over R_2}\) 
\overline{\l_1}^{(m,n)}\l_2^{(m,n)} + c.c. \)\;.
\eea
This is the kinetic term for the Dirac fermion $\l_D = (\l_1,\l_2)$ with mass
$M(m,n)$. Together with the vector $V_\m^{(m,n)}$ and the scalars 
$\Pi_{1,2}^{(m,n)}$, $\l_D$ forms a massive $N=1$ vector multiplet in 4d.

So far the massless sector of the theory has an unwanted N=2 supersymmetry.
This can be reduced
by considering instead of the torus $T^2$ the orbifold $T^2/Z_2$. Under
the corresponding reflection $(y,z)\rightarrow (-y,-z)$ vectors and scalars
are even and odd, respectively,
\bea\label{z1v}
PV_\m(x,-y,-z)P^{-1} = +V_\m(x,y,z)\;,\quad 
PV_{5,6}(x,-y,-z)P^{-1} =-V_{5,6}(x,y,z)\;,
\eea
where we choose $P=I$. This implies for the Kaluza-Klein modes,
\bea
V_\m^{(-m,-n)} &=& +V_\m^{(m,n)} = +V_\m^{(m,n)\dagger}\;,\\
V_{5,6}^{(-m,-n)} &=& -V_{5,6}^{(m,n)} = +V_{5,6}^{(m,n)\dagger}\;.
\eea
In this way scalar zero modes are obviously eliminated. Further, the number
of massive KK modes is halved. Since the derivatives $\partial_{5,6}$ are odd
under reflection the two Weyl fermions $\l_1$ and $\l_2$ must have opposite
parities,
\bea\label{z1l}
P\l_1(x,-y,-z)P^{-1} = + \l_1(x,y,z)\;,\quad
P\l_2(x,-y,-z)P^{-1} = - \l_2(x,y,z)\;.
\eea
Comparison of eqs.~(\ref{z1v}) and (\ref{z1l}) shows that 
$(V_\m,\l_1)$ and $(V_{5,6},\l_2)$
form vector and chiral multiplets, respectively. Only vector
multiplets have zero modes. The orbifold compactification breaks the
extended supersymmetry which one obtains from the 6d theory by dimensional
reduction. This is completely analogous to the previously discussed 
compactification of 5d theories on $S^1/Z_2$ \cite{mp98}.

The zero modes obtained by compactification on the orbifold $T^2/Z_2$
form a $N=1$ supersymmetric SO(10) theory in 4d. A breaking of the full
SO(10) gauge group can be achieved by using the two parities $P_{GG}$ and
$P_{PS}$ which define the symmetric subgroups 
G$_{GG}$=SU(5)$\times$U(1)
and G$_{PS}$=SU(4)$\times$SU(2)$\times$SU(2), respectively. In the vector
representation the parities can be taken as ($\s_0=I$),
\bea
P_{GG} = \(\begin{tabular}{ccccc} $\s_2$&0&0&0&0 \\ 0&$\s_2$&0&0&0 \\
0&0&$\s_2$&0&0 \\ 0&0&0&$\s_2$&0 \\ 0&0&0&0&$\s_2$ \end{tabular}\)\;,\quad
P_{PS} = \(\begin{tabular}{ccccc} $-\s_0$&0&0&0&0 \\ 0&$-\s_0$&0&0&0 \\
0&0&$-\s_0$&0&0 \\ 0&0&0&$\s_0$&0 \\ 0&0&0&0&$\s_0$ \end{tabular}\)\;.
\label{PGG-PS}
\eea
\begin{table}[t]
\begin{center}
\renewcommand{\arraystretch}{1.5}
\begin{tabular}{|c|c|c||ccc||ccc|}
 \hline
 &  &  &
 \multicolumn{3}{ c||}{($V_\mu$,~$\lambda_1$)} &
 \multicolumn{3}{ c|}{($V_{5,6}$,~$\lambda_2$)}
 \\
 \cline{4-9}
 G$_{SM'}$ &
 G$_{GG}$  &
 G$_{PS}$ &
 $Z_2$ &
 $Z_2^{GG}$ &
 $Z_2^{PS}$ &
 $Z_2$ &
 $Z_2^{GG}$ &
 $Z_2^{PS}$
 \\
 \hline
  \itshape \bfseries
  ( 8, 1, 0, 0) & (24, 0) &  (15,1,1)
   & + & + & + & $-$ & $-$ & $-$
 \\
  ( 3, 2,$-$5, 0) & (24, 0) & (6,2,2)
   & + & + & $-$& $-$ & $-$ & +
 \\
  ( $\overline{3}$, 2, 5, 0) & (24,0) & (6,2,2)
   & + & + & $-$ & $-$ & $-$ & +
 \\
  \itshape \bfseries
  ( 1, 3, 0, 0) & (24,0) & (1,3,1)
   & + & + & + & $-$ & $-$ & $-$
 \\
  \itshape \bfseries
  ( 1, 1, 0, 0) & (24,0) & (1,1,3)
   & + & + & + & $-$ & $-$ & $-$
 \\
  ( 3, 2, 1, 4) & (10,4) & (6,2,2)
   & + & $-$ & $-$ & $-$ & + & +
 \\
  ( $\overline{3}$, 1, $-4$, 4) & (10,4) & (15,1,1)
   & + & $-$ & + & $-$ & + & $-$
 \\
  ( 1, 1, 6, 4) & (10,4) & (1,1,3)
   & + & $-$ & + & $-$ & + & $-$
 \\
  ( $\overline{3}$, 2, $-1$, $-4$) & ($\overline{10}$,$-$4) & (6,2,2)
   & + & $-$ & $-$ & $-$ & + & +
 \\
  ( 3, 1, 4, $-4$) & ($\overline{10}$,$-$4)  &  (15,1,1)
   & + & $-$ & + & $-$ & + & $-$
 \\
  ( 1, 1, $-6$, $-4$) & ($\overline{10}$,$-$4) &  (1,1,3)
   & + & $-$ & + & $-$ & + & $-$
 \\
  \itshape \bfseries
  ( 1, 1, 0, 0) & (1,0) & (15, 1, 1)
   & + & + & + & $-$ & $-$ & $-$
 \\
 \hline
\end{tabular}
\end{center}
\caption{Parity assignment for the components 
$V_M^A = \frac{1}{2} \tr (T^A V_M) $ of the $\bf 45$-plet of SO(10).
G$_{SM'}$ = SU(3)$\times$SU(2)$\times$U(1)$^2$,
G$_{GG}$ = SU(5)$\times$U(1) and G$_{PS}$ = SU(4)$\times$SU(2)$\times$SU(2).}
\label{tab_adj}
\end{table}
For the vector fields and the gauginos $\l_1$ one demands
\bea
P_{GG}V_\m(x,-y,-z+{\p R_2/2})P_{GG}^{-1} &=& 
+ V_\m(x,y,z+{\p R_2/2})\;, \NO\\
P_{PS}V_\m(x,-y+{\p R_1/2},-z)P_{PS}^{-1} &=& 
+ V_\m(x,y+{\p R_1/2},z)\;.
\eea 
Component fields belonging to the symmetric subgroup G$_s$ 
then have positive parity, those of the coset space SO(10)/G$_s$
have negative parity. The restrictions of the discrete symmetry 
$Z_2$ require the opposite parities for the
scalars and the gauginos $\l_2$, 
\bea
P_{GG}V_{5,6}(x,-y,-z+{\p R_2/2})P_{GG}^{-1} &=& 
-V_{5,6}(x,y,z+{\p R_2 /2})\;,\NO\\
P_{PS}V_{5,6}(x,-y+{\p R_1/2},-z)P_{PS}^{-1} &=&
-V_{5,6}(x,y+{\p R_1/2},z)\;.
\eea 
The component fields are again split. Their parities
are given in table~\ref{tab_adj} for the different G$_{SM'}$
representations contained in the $\bf 45$-plet of SO(10). 

The mode expansion for the fields $\Phi(x,y,z)$ with any combination of 
parities reads explicitly, 
\bea
\Phi_{+++}(x,y,z) &=& {1\over \p \sqrt{R_1R_2}} \sum_{m \ge 0} 
{1\over 2^{\d_{m,0}\d_{n,0}}} \phi_{+++}^{(2m,2n)}(x)
\cos\({2my\over R_1}+{2nz\over R_2}\) ,\\
\Phi_{++-}(x,y,z) &=& {1\over \p \sqrt{R_1R_2}} \sum_{m \ge 0}
\phi_{++-}^{(2m,2n+1)}(x)
\cos\({2my\over R_1}+{(2n+1)z\over R_2}\) ,\\
\Phi_{+-+}(x,y,z) &=& {1\over \p \sqrt{R_1R_2}} \sum_{m \ge 0}
\phi_{+-+}^{(2m+1,2n)}(x)
\cos\({(2m+1)y\over R_1}+{2nz\over R_2}\) ,\\
\Phi_{+--}(x,y,z) &=& {1\over \p \sqrt{R_1R_2}} \sum_{m \ge 0}
\phi_{+--}^{(2m+1,2n+1)}(x)
\cos\({(2m+1)y\over R_1}+{(2n+1)z\over R_2}\) ,\\  
\Phi_{-++}(x,y,z) &=& {1\over \p \sqrt{R_1R_2}} \sum_{m \ge 0}
\phi_{-++}^{(2m+1,2n+1)}(x)
\sin\({(2m+1)y\over R_1}+{(2n+1)z\over R_2}\) ,\\ 
\Phi_{-+-}(x,y,z) &=& {1\over \p \sqrt{R_1R_2}} \sum_{m \ge 0}
\phi_{-+-}^{(2m+1,2n)}(x)
\sin\({(2m+1)y\over R_1}+{2nz\over R_2}\) ,\\  
\Phi_{--+}(x,y,z) &=& {1\over \p \sqrt{R_1R_2}} \sum_{m \ge 0}
\phi_{-+-}^{(2m,2n+1)}(x)
\sin\({2my\over R_1}+{(2n+1)z\over R_2}\) ,\\ 
\Phi_{---}(x,y,z) &=& {1\over \p \sqrt{R_1R_2}} \sum_{m \ge 0}
\phi_{---}^{(2m,2n)}(x)
\sin\({2my\over R_1}+{2nz\over R_2}\) .
\eea
Only fields for which
all parities are positive have zero modes; they form an $N=1$ massless
vector multiplet in the adjoint representation of the unbroken group
G$_{SM'}$. All other fields with one or more negative parity combine
to massive vector multiplets with some G$_{SM'}$ quantum numbers.

It is interesting to consider the two limiting cases $R_1 \rightarrow 
0$ with $R_2$ fixed, and $R_1$ fixed with  $R_2 \rightarrow 0$.  
The dependence on one of the compact dimensions then disappears and
one considers effectively a 5 dimensional subspace.
In the first case SO(10) is broken to 
the Pati-Salam group G$_{PS}$ = SU(4)$\times$SU(2)$\times$SU(2),
in the second case it is broken to
the extended Georgi-Glashow group G$_{GG}$ = SU(5)$\times$U(1). Only
for finite $R_1$ and $R_2$ one obtains the extended standard model group
G$_{SM'}$ = SU(3)$\times$SU(2)$\times$U(1)$^2$.

Matter fields are easily added to the 6d supersymmetric Yang-Mills theory. Of 
particular interest is the $\bf 10$-plet of `Higgs' fields. It contains
two complex scalars $H$ and $H'$, and a fermion $h = (h,h')$ where
$h$ and $h'$ have 4d chiralities $\g_5 h = h$, $\g_5 h' =-h'$, and
positive 6d chirality $\G_7 h = h$.
The corresponding lagrangian reads \cite{fay85}
\bea
{\cal L}^M_{6d} &=& |D_M H|^2 + |D_M H'|^2
 -{1\over 2} g^2 \(H^{\dagger}T^A H + H'^{\dagger}T^A H'\)^2\NO\\ 
&&+i\overline{h}\G^M D_M h  
-i\sqrt{2}g \(\overline{h}\L H + \overline{h}\L^c H' + c.c.\)\;.
\eea
\begin{table}[t]
\begin{center}
\renewcommand{\arraystretch}{1.5}
\begin{tabular}{|c|c|c||ccc||ccc|}
 \hline
  &   &  &
 \multicolumn{3}{ c||}{($H_1,h_1)$} &
 \multicolumn{3}{ c|}{($H_2,h_2)$}
 \\
 \cline{4-9}
 G$_{SM'}$ &  $G_{GG}$ & $G_{PS}$ &
 $Z_2$ &
 $Z_2^{GG}$ &
 $Z_2^{PS}$ &
 $Z_2$ &
 $Z_2^{GG}$ &
 $Z_2^{PS}$
 \\
 \hline
   ( 3, 1, $-$2, 2) & (5,2) & (6,1,1)
   & + & + & $-$ & $+$ & $-$ & $-$
 \\
  \itshape \bfseries
   ( 1, 2, 3, 2) & (5,2) & (1,2,2)
   & + & + & + & + & $-$ & +
 \\
  ( $\overline{3}$, 1, 2, $-$2) & ($\overline{5}$,$-$2)  &  (6,1,1)
   & + & $-$ & $-$ & + & + & $-$
 \\
  \itshape \bfseries
  ( 1, 2, $-$3, $-$2) & ($\overline{5}$,$-$2)  & (1,2,2)
   & + & $-$ & + & + & + & +
 \\
 \hline
\end{tabular}
\end{center}
\caption{Parity assignment for the components of the Higgs 
$\bf 10$-plets of SO(10).
The parities of $H_{1,2}'$ and $h_{1,2}'$ are opposite to the listed parities
of $H_{1,2}$ and $h_{1,2}$.
G$_{SM'}$ = SU(3)$\times$SU(2)$\times$U(1)$^2$,
G$_{GG}$ = SU(5)$\times$U(1) and G$_{PS}$ = SU(4)$\times$SU(2)$\times$SU(2).}
\label{tab_hig}
\end{table}
After integrating over the compact dimensions one obtains for the kinetic
terms of the 4d fields,
\bea
{\cal L}_4^{(3)} &=& \sum_{m,n} \(
i\overline{h}^{(m,n)}\g^\m\partial_\m h^{(m,n)} +
i\overline{h'}^{(m,n)}\g^\m\partial_\m h'^{(m,n)}  \right.\NO\\
&& \vspace{-.2cm}\hspace{1cm} + \({m\over R_1} -i{n\over R_2}\) 
\overline{h}^{(m,n)}h'^{(m,n)} + c.c.  \NO\\
&&\hspace{1cm}+\partial_\m H^{(m,n)\dagger}\partial^\m H^{(m,n)}
+ M(m,n)^2 H^{(m,n)\dagger}H^{(m,n)} \NO\\
&&\hspace{1cm}\left.+\partial_\m H'^{(m,n)\dagger}\partial^\m H'^{(m,n)}
+ M(m,n)^2 H'^{(m,n)\dagger}H'^{(m,n)}\)\;.
\eea
We now have to define the action of the parities on the Higgs fields. 
For the N=1 supermultiplets $H=(H,h)$ and $H'=(H',h')$ we have opposite
parities with respect to every $Z_2$. Note, however, that there is an
ambiguity in the global sign of each parity, as long as we do not consider
a superpotential for the matter fields.

For the $Z_2$ parity we choose
\bea
P H(x,-y,-z) &=& + H(x,y,z)\;,  \\
P H'(x,-y,-z) &=& - H'(x,y,z)\;, 
\eea
with $P=I$. This breaks the extended supersymmetry as in the case of the
$\bf 45$-plet. 
For $Z_2^{GG}$ we can take
\bea
P_{GG}H(x,-y,-z+\p R_2/2) &=&  + H(x,y,z+\p R_2/2)\;,\\
P_{GG}H'(x,-y,-z+\p R_2/2) &=& - H'(x,y,z+\p R_2/2)\;,
\eea
where the matrix representations of $P_{GG}$ is 
defined in eq.~(\ref{PGG-PS}). Note that the SU(5) 
$\bf 5$- and $\bf 5^*$-plets contained in the 
$\bf 10$-plet have opposite parities with respect to $Z_2^{GG}$ 
(cf.~table~\ref{tab_hig}).
The parity $P_{PS}$ yields the desired doublet-triplet splitting. 
As mentioned above, it is ambiguous up to a sign, like Kawamura's 
parity for the SU(5) GUT \cite{kaw00}. With 
\bea
P_{PS}H(x,-y+\p R_1/2,-z) &=& +H(x,y+\p R_1/2,z)\;,\\
P_{PS}H'(x,-y+\p R_1/2,-z) &=& -H'(x,y+\p R_1/2,z)\;,  
\eea
one obtains one SU(2) doublet N=1 supermultiplet as zero modes, 
as given in table~\ref{tab_hig}. The related SU(3) triplet
is heavy. The opposite choice of sign leads to a massless
colour triplet and a heavy weak doublet. 

In order to obtain the wanted two Higgs doublets as zero modes one has 
to introduce two $\bf 10$-plets $H_1$ and $H_2$. Their parities must
be different with respect to $Z_2^{GG}$, as listed in table~\ref{tab_hig}.
Note, that the irreducible 6d gauge anomalies of the two $\bf 10$-plets
cancel the one of the $\bf 45$-plet \cite{hmr01}.

So far we have only considered the breaking of the SO(10) GUT symmetry.
Quarks and leptons can be included along the lines discussed for the 5d
SU(5) GUTs \cite{afx01,hmr01}, e.g. as $\bf 16$-plets on a brane.
The electroweak gauge group SU(2)$\times$U(1)$^2$, which contains
$U(1)_{B-L}$, is the minimal extension of the chiral standard model 
electroweak symmetry which can keep all fermions, including the right-handed 
neutrino, massless. Its breaking is related to the generation of fermion
masses which will be discussed elsewhere. \\

\noindent
{\bf Acknowledgement}\\
\noindent
We would like to thank Arthur Hebecker and Jan Louis for helpful discussions.



\begin{thebibliography}{99}

\bibitem{gfm75}
H.~Georgi, Particles and Fields 1974, ed. C.~E.~Carlson (AIP, NY, 1975) 
p. 575;\\
H.~Fritzsch, P.~Minkowski, \ap{93}{1975}{193}

\bibitem{ps74}
J.~C.~Pati, A.~Salam, \pr{10}{1974}{275}

\bibitem{gg74}
H.~Georgi, S.~L.~Glashow, \prl{32}{1974}{438}

\bibitem{kaw00}
Y.~Kawamura, \ptp{103}{2000}{613};\ {ibid.}{\bf 105} (2001) 691

\bibitem{afx01}
G.~Altarelli, F.~Feruglio, \pl{511}{2001}{257};\\
A.~B.~Kobakhidze, \pl{514}{2001}{131};\\
L.~J.~Hall, Y.~Nomura, hep-ph/0103125;\\
T.~Kawamoto, Y.~Kawamura, hep-ph/0106163;\\
A.~Hebecker, J.~March-Russell, hep-ph/0106166;\\
R.~Barbieri, L.~J.~Hall, Y.~Nomura, hep-ph/0106190; hep-ph/0107004;\\
A.~E.~Faraggi, hep-ph/0107094;\\
N.~Haba, Y.~Shimizu, T.~Suzuki, K.~Ukai, hep-ph/0107190;\\
T.~Li, hep-th/0107136;\\ 
L.~J.~Hall, H.~Murayama, Y.~Nomura, hep-th/0107245

\bibitem{hmr01}
A.~Hebecker, J.~March-Russell, hep-ph/0107039

\bibitem{sla81}
R.~Slansky, \prep{79}{1981}{1}

\bibitem{fay85}
P.~Fayet, \pl{159}{1985}{121}

\bibitem{mp98}
E.~A.~Mirabelli, M.~E.~Peskin, \pr{58}{1998}{065002}


\end{thebibliography}
\end{document}